\documentclass[conference]{IEEEtran}

\usepackage{amsmath,amssymb,amsfonts}
\usepackage{algorithmic}
\usepackage{graphicx}
\usepackage{textcomp}
\usepackage{xcolor}
\usepackage{hyperref}
\usepackage{makecell}
\usepackage{booktabs}
\usepackage{ifthen}
\usepackage{array}
\usepackage{biblatex}
\usepackage{datetime2}

\def\BibTeX{{\rm B\kern-.05em{\sc i\kern-.025em b}\kern-.08em
    T\kern-.1667em\lower.7ex\hbox{E}\kern-.125emX}}

\begin{document}

\pagenumbering{arabic}
\pagestyle{plain} 

\title{What's So Hard about the Monty Hall Problem?}

\author{
\IEEEauthorblockN{Rafael C. Alvarado}
\IEEEauthorblockA{School of Data Science \\
University of Virginia \\
Charlottesville, VA USA \\
\today \\
https://orcid.org/0000-0002-7218-0114}}

\maketitle

\begin{abstract}
The Monty Hall problem is notorious for its deceptive simplicity. Although today it is widely used as a provocative thought experiment to introduce Bayesian thinking to students of probability, in the not so distant past it was rejected by established mathematicians. This essay provides some historical background to the problem and explains why it is considered so counter-intuitive to many. It is argued that the main barrier to understanding the problem is the back-grounding of the concept of dependence in probability theory as it is commonly taught. To demonstrate this, a Bayesian solution is provided and augmented with a probabilistic graphical model (PGM) inspired by the work of Pearl (1988, 1998) \cite{pearlProbabilisticReasoningIntelligent1988}\cite{pearl1998graphical}. Although the Bayesian approach produces the correct answer, without a representation of the dependency structure of events implied by the problem, the salient fact that motivates the problem's solution remains hidden.
\end{abstract}


\section{Introduction}

Every student of probability has heard of the Monty Hall problem. Notorious for its counter-intuitive solution, when it was first posed in a letter to the editor in \textit{The American Statistician} (Selvin 1975: 67) \cite{selvinLettersEditor1975} and later in \textit{Parade} magazine (vos Savant 2018 [1990]), its solution was rejected, often vehemently, by a majority of respondents, many of whom should have known better. Even Paul Erdős, one of the greatest mathematicians of the twentieth century, was skeptical of its solution. He changed his mind only after being shown the correct answer through an empirical demonstration (Vazsonyi 2002: 5) \cite{vazsonyiWhichDoorHas2002}. Mathematicians are not fond of existence proofs. Although introduced as an entertaining brainteaser, the problem has been more like the Riddle of the Sphinx, exposing hubris as much as knowledge in those who seek to answer it.\footnote{For good examples of hubris, in the form of severe mansplaining, see "The Time Everyone `Corrected' the World’s Smartest Woman" (Crocket 2015) \cite{crockettTimeEveryoneCorrected2015} and "Game Show Problem" (vos Savant 1991) \cite{vossavantGameShowProblem2018}.}

\subsection{The Problem}

For those not familiar with the problem, it goes like this. Imagine you are a contestant on \textit{Let’s Make a Deal}, a famous American game show hosted by Monty Hall. In the event, the host presents you with three closed doors---let’s call them $A$, $B$, and $C$---and informs you that behind one of the doors is a brand new car, while behind the two others are goats. You really want the car, and if you can guess the door the car is behind, you get to keep it. The host invites you to guess which door contains the car and so you guess door $A$. At this point, instead of letting you know if you guessed correctly, the host, who knows where the car is, opens a door that you did not guess, door $B$, which reveals a goat. The host then gives you the opportunity to change your guess to the remaining door, door $C$. The question is, should you stay with door $A$ or switch to door $C$?

Of course, there are many versions of the problem---Selvin’s involves three boxes and a set of keys, for example---but they are all essentially the same, although some are more confusing than others. One point of confusion that is hard to eliminate, pointed out by the \textit{The Angry Statistician}, is the ambiguity regarding the host’s \textit{modus operandi}---does he reveal a door and offer you to switch based on your choice? For example, does he open a door and invite you to switch only when you have guessed correctly the first time? Here, as with the great majority of responses, we shall assume that he does not do anything of the sort, and that his action to reveal a goat happens either each time the game is played or randomly. This is a big assumption, but it is not unreasonable. If we do not make it, the problem becomes a trick question, as there is no way to know if the host does this or not---although, as we shall see, we can easily model this possibility.

\subsection{The Solution}

Now, the surprising solution to the question of whether you should keep door $A$ or switch to door $C$ is that you should switch. If you stick with door $A$, the probability of winning the car is $\frac{1}{3}$. If you switch to door $C$, it is $\frac{2}{3}$. 

This conclusion can be demonstrated both experimentally and logically. Experimentally, it is easy to reproduce the conditions of the game show in real life and tally the results after a sufficient number of trials.\footnote{You can find a collection of computer simulations to do this for you, in a variety of languages, here: \url{https://rosettacode.org/wiki/Monty_Hall_problem}.} Logically, the solution can be demonstrated by describing the sample space of the game—the list of all possible outcomes where the contestant decides to switch—and then adding up the results associated with each possibility (see Table \ref{tab:sample-space-01}, based on Selvin’s original letter).

\begin{table}
    \centering
    \begin{tabular}{ccccc} 
    \toprule
     \makecell{Car\\ behind}& \makecell{Contestant\\ chooses}& \makecell{Monty Hall\\ opens}& \makecell{Contestant\\ switches to}& Result \\ 
     \midrule
    A& A& B  or C& C or B& L\\
    A & B & C & A & W\\  
    A & C & B & A & W\\ 
    \midrule
 B& A& C& B&W\\ 
 B& B& A or C& C or A&L\\ 
 B& C& A& B&W\\ 
 \midrule
 C& A& B& C&W\\
 C& B& A& C&W\\
 C& C& A or B& B or A&L\\
 \bottomrule
    \end{tabular}
    \vspace{5px}
    \caption{Sample space, based on Colvin 2014.}
    \label{tab:sample-space-01}
\end{table}

In Table \ref{tab:sample-space-01}, each row corresponds to a possible outcome of the game, each of which has the same probability, $\frac{1}{9}$. If you add up the winning rows, you get $\frac{6}{9}$, or $\frac{2}{3}$; the losing rows add up to $\frac{3}{9}$, or $\frac{1}{3}$. Note that the first, middle, and last rows actually define two outcomes each, but the probabilities of each of these is half of $\frac{1}{3}$, or $\frac{1}{6}$, so the sums are the same. But to point this out is to get ahead of ourselves.

\subsection{The Response}

These results, plain as they are, come as a surprise because they conflict with the obvious answer one arrives at by applying what seem to be basic principles of probability theory. In this approach, it does not matter if you switch, since both remaining doors have the same probability of hiding the car. The reasoning is as follows. In the first guess, the contestant has three doors to choose from, and so, given no other information about the situation, each door has a $\frac{1}{3}$ chance of hiding the car. Once the host opens the second door, though, there are only two doors left, and so, \textit{given no other information about the situation}, each door has a $\frac{1}{2}$ chance of hiding the car. The probability is simply a matter of evenly dividing the probability over the number of available doors. Yes, the probability of the remaining door is increased, but so is that of the door first guessed, and so the contestant gains nothing by switching. 

The problem with this answer is contained in the assumption that, after the host opens a door, there is no information to tip the balance in favor of the two remaining doors. Actually, there is new information available, but it is hard to see. The reason is our entrenched bias in favor  the independence assumption---the idea that events are independent until proven to be otherwise. In contrast to this assumption, the answer to the Monty hall problem hinges precisely on understanding the dependent relationship between two events---the initial choice by the contestant, and the second choice prompted by the hosts action.

In fairness, one can understand the resistance to accept the correct explanation as given. After all, the logical and empirical solutions are both merely inductive, not deductive, as the second approach is. They arrive at the solution through brute force, by generating a set of results and counting wins and losses, and so are unsatisfying. Such approaches also invite the nagging suspicion, unfounded as it may be, that the generative mechanisms behind them are flawed in some way, and may be producing the wrong results. Consider, for example, if we represented the sample space in Table \ref{tab:sample-space-01} by breaking out the lines we collapsed---the first, middle, and last lines, where the host has a choice of doors to open. If we tally up the results with the assumption that each row in the new table (Table \ref{tab:expanded-sample-space}) has the same probability---an easy mistake to make---we get an equal tally for wins and losses.

\begin{table}
    \centering
    \begin{tabular}{ccccc}
    \toprule
    \makecell{Car\\behind}& \makecell{Contestant\\chooses} & \makecell{Monty Hall\\opens} & \makecell{Contestant\\switches to} & Result\\
    \midrule
    A & A & B& C&L\\
    A & A & C& B&L\\
    A & B & C& A&W\\
    A & C & B& A&W\\
    \midrule
    B & A & C& B&W\\
    B & B & A& C&L\\
    B & B & C& A&L\\
    B & C & A & B&W\\
    \midrule
    C & A & B& C&W\\
    C & B & A& C&W\\
    C & C & A& B&L\\
    C & C & B& A&L\\
    \bottomrule
    \end{tabular}
    \vspace{5px}
    \caption{Sample space with expanded possibilities}
    \label{tab:expanded-sample-space}
\end{table}

Of course, to correct this error, we just need to assign probabilities to each event in the sample space, but this moves us beyond the original simplicity of the inductive approach and the clarity of its tabular demonstration. Table \ref{tab:expanded-sample-space-with-probs} begs for a more formal approach that would eliminate its redundancy and explain the values provided.

\begin{table}
    \centering
    \begin{tabular}{cccccc}
    \toprule
    \makecell{Car\\behind}& \makecell{Contestant\\chooses} & \makecell{Monty Hall\\opens} & \makecell{Contestant\\switches to} & Result & P\\
    \midrule
    A & A & B& C&L& $\frac{1}{18}$ \\
    A & A & C& B&L& $\frac{1}{18}$ \\
    A & B & C& A&W& $\frac{1}{9}$ \\
    A & C & B& A&W& $\frac{1}{9}$ \\
    \midrule
    B & A & C& B&W& $\frac{1}{9}$ \\
    B & B & A& C&L& $\frac{1}{18}$ \\
    B & B & C& A&L& $\frac{1}{18}$ \\
    B & C & A & B&W& $\frac{1}{9}$ \\
    \midrule
    C & A & B& C&W& $\frac{1}{9}$ \\
    C & B & A& C&W& $\frac{1}{9}$ \\
    C & C & A& B&L& $\frac{1}{18}$ \\
    C & C & B& A&L& $\frac{1}{18}$ \\
    \bottomrule
    \end{tabular}
    \vspace{5px}
    \caption{Sample space with probabilities}
    \label{tab:expanded-sample-space-with-probs}
\end{table}

\section{The Bayesian Approach}

Because of the importance of grasping the dependency between events in the correct answer, a Bayesian approach is often employed to provide a formal solution. Unlike classical probability theory, Bayesian probability foregrounds the relationship between events through the concept of conditional probability. 

\subsection{The Standard Solution}

In its canonical form, which is easily derived from the axioms of probability, the theorem looks like this:

$$
P(H|e) = \frac{P(H)P(e|H)}{P(e)}
$$

In this formulation, $H$ is the hypothesis and $e$ the evidence. The formula is used when we want to know how probable an hypothesis is for explaining, or accounting for, the known existence of some evidence, and when we know, or can provide good estimates of, things on the right side of the formula. A common example of an hypothesis and evidence is a disease, such as small pox, and a symptom, such as red spots on the skin.  Traditionally, the terms of this formula are given names: $P(H|e)$ is called the \textit{posterior} probability of $H$ (given $e$), $P(H)$ is the \textit{prior} probability of $H$, $P(e|H)$ is the \textit{likelihood} of $e$ (given $H$), and $P(e)$ is the \textit{marginal} probability of $e$.\footnote{Here it is not necessary to go into depth about Bayes’ theorem, other than to show how it can be applied to the Monty Hall problem. For a lucid and concise introduction to the subject, see James Stone’s \textit{Bayes’ Rule: A Tutorial Introduction to Bayesian Analysis} (2013) \cite{stoneBayesRuleTutorial2013}.}

When applied to the Monty Hall problem, the Bayesian approach proceeds by plugging in the proper values to the theorem and then computing the results. In our example, the two possible locations of the car---behind doors $A$ and $C$---are considered to be hypotheses (the only two possible), while the door opened by the host, door $B$, is the evidence. So, if we let $Car_A$ and $Car_C$ stand for the hypotheses that the car is behind doors $A$ and $C$ respectively, and also let $Host_B$ stand for the host’s revelation that door $B$ contains a goat, we get the following equations to solve:

$$
Switch: P(Car_C|Host_B) = \frac{P(Car_C)P(Host_B|Car_C)}{P(Host_B)}
$$

and

$$
Keep: P(Car_A|Host_B) = \frac{P(Car_A)P(Host_B|Car_A)}{P(Host_B)}
$$

The first formula corresponds to the decision to switch doors and reads: the probability that the car is behind door $C$, in the case where the host opened door $B$ (and the contestant guessed door A), is equal to the probability that the car is behind door $C$ in any case, times the probability that the host would open door $B$ in the case that the car is behind door $C$---remember, he knows where the car is---divided by the probability that the host would open door B in any case. (Here, I have used the phrase “in the case that” instead of the usual “given that,” or the more obscure “conditioned on,” to clarify the logic.)  The second formula represents the contestant's decision to turn down the host's offer to switch, and defines the probability that the car is behind door $A$, the original choice.

Now, since we want to compare the two hypotheses, we can apply the odds ratio version of Bayes’ rule, eliminating the need to calculate the marginal $P(Host_B)$:

$$
\frac{P(Car_C)P(Host_B|Car_C)}{P(Car_A)P(Host_B|Car_A)}
$$

To calculate the value of this expression, we just need to calculate the values of each probability within it. These are as follows. First, we know the prior probabilities

$$
P(Car_A) = P(Car_C) = \frac{1}{3}
$$

since the probability that the car is behind any door, without taking into account any information provided by the host’s opening of a door, is $\frac{1}{3}$. Since the two priors are the same, we can remove them and just worry about the ratio of likelihoods:

$$
\frac{P(Host_B|Car_C)}{P(Host_B|Car_A)}
$$

Now, to calculate these, we have to do a little close reading of our problem. In effect, these likelihoods concern the choices available to the host based on his knowledge of both the location of the car and the door guessed by the contestant. In other words, it is in the likelihoods where we may observe the information that is conveyed by the host in making his decision. 

Earlier we noted that the incorrect deduction of the answer using basic probability theory is wrong because it assumes that the event $Host_B$ is independent of either $Car_C$ or $Car_A$, i.e. it assumed that $P(Host_B|Car_C) = P(Host_B|Car_A) = P(Host_B)$. In fact, the host's decision regarding which door to open is highly dependent on which door the contestant has chosen. In the case where the contestant guesses correctly---the case of $Car_A$---the host has two choices of door to open, and so $P(Host_B|Car_A)$ is $\frac{1}{2}$, since he could have also opened door $C$. However, in the case where the contestant guessed \textit{incorrectly}---the case of $Car_C$---the host has only one choice, and so $P(Host_B|Car_C)$ is $1$. Given this, we get the following ratios:

$$
\frac{P(Host_B|Car_C)}{P(Host_B|Car_A)}
= \frac{1}{0.5}
= \frac{2}{1}
= \frac{W}{L}
$$

Thus, we see that if the contestant switches and chooses car $C$, the odds of winning are $2:1$, i.e. winning by switching has a probability of $\frac{2}{3}$.

\subsection{Limitations of the Approach}

The advantage of the Bayesian approach is that it arrives at the correct answer by means of deduction and is therefore more satisfying mathematically. It also has the virtue of demonstrating the utility of Bayes’ theorem, which has until relatively recently been the object of great suspicion by traditional statisticians (see McGrayne 2011) \cite{mcgrayneTheoryThatWould2012}. However, the problem with this approach is that, like the other methods described here, it hides the logic of the solution to the problem behind the artifacts of an algebraic process. To use an idiom from journalism, such solutions bury the lede, the most important part of the story, which is that Monty Hall is constrained by a specific set of rules to such a degree that he can be replaced by a machine, a fact already implied by our ability to simulate the game programmatically. In each formulation of the correct solution, these rules are alluded to---as by the middle column of the table to the first solution and in the calculation of the likelihood in the Bayesian one---but they are never made explicit. This is a shame because, once this description is given, it’s not only easy to understand the solution intuitively, it is possible to see that there are other solutions as well.

\section{Dependency Structures}

The Monty Hall Problem is hard because the most important element of its solution, the dependency structure that constrains the host’s actions, is never foregrounded and given its proper place. In the variant retellings and clarifications of the problem, the role of the host has remained Oz-like, a man behind a curtain whose projected image is much larger and frightening than his real person. As a result, the solution to the problem has often been explained as if the host’s intentions and motives somehow matter. They do not—at least as long as we make the reasonable assumption that the problem is being posed in good faith, and not playing on the ambiguity we noted earlier, regarding the host’s action to reveal a door with a goat. 

\subsection{Two Scenarios}

Let’s pull back the curtain and scrutinize the Bayesian solution to clarify the role of the host. To do this, we may summarize the results of the Bayesian approach in the following way.\footnote{The description provided here is very close to that provided by Leonard Mlodinow in \textit{The Drunkard’s Walk} (2009) \cite{mlodinowDrunkardWalkHow2008}.} Although the full sample space of the problem contains every combination of choices by the host in placing the car and revealing a goat, and contestant in guessing and then switching or not, there are actually only two relevant \textit{scenarios} of the game, and these depend entirely on a comparison of where the car is placed at the outset and what the contestant guessed first. Both of these facts the host knows and must consider in making his decision of which door to open.  

The first scenario is that the contestant guesses correctly the first time. The second is that the contestant guesses incorrectly the first time. 

The first scenario has a probability of $\frac{1}{3}$, while the second has a probability of $\frac{2}{3}$, since there are two incorrect doors and one correct door to choose from.  Obviously, if the contestant is right and switches, she will lose. But, if the contestant is wrong and switches, she will win. In this scenario, if a door is revealed to be empty, the other door is the only one remaining and so must have the car---\textit{the host has no choice but to reveal the door that does not hide the car}. Therefore, the contestant has a $\frac{2}{3}$ chance of winning if they switch. So, she should switch.\footnote{Put more formally, when the contestant makes her first guess, we know that the door she guesses has a $\frac{1}{3}$ chance of being correct. It therefore follows that the other two doors, considered together, have a $\frac{2}{3}$ chance of being correct—in other words, if $P(Car_A) = \frac{1}{3}$ then $P(\neg Car_A) = \frac{2}{3}$, and since $\neg Car_A = Car_B \| Car_C$, then $P(Car_B \| Car_C) = \frac{2}{3}$. Now, when the host opens door $B$, $P(Car_A)$ is still $\frac{1}{3}$ and thus $P(Car_B \| Car_C)$ is still $\frac{2}{3}$, but since now we know that $P(Car_B) = 0$, it follows that $P(Car_C) = \frac{2}{3}$. }

Again, the host’s decision about which door to reveal depends on the outcomes of two prior events---the placing of the car and the contestant’s first guess.  The two possible combinations of these events determine the entire outcome. 

In the application of Bayes’ theorem this salient fact is not obvious. It is back-grounded by the formula and appears only in our verbal discussion of the two likelihoods, where we go off-line, as it were, to determine the probabilities of the host’s decision in each case. To be sure, the formula does lead us to that discussion and, of course, it does produce the correct solution. But by itself it does not provide an answer that satisfies our desire for a solution that fully explains the why behind the answer.

A complete and satisfying representation of the problem would include a description of all the relevant events along with a description of how each event depends on the others, to the extent that they do. In describing these dependencies, it would include information not simply about which events influence others, but specifically how they do, to the extent that this information is available.

\subsection{Probabilistic Graphic Models}

Fortunately, we have a tool for doing precisely this: probabilistic graphical models, or PGMs (see Pearl 1988, 1998) \cite{pearlProbabilisticReasoningIntelligent1988}\cite{pearl1998graphical}. A PGM is a kind of graph, or network, that contains a set of vertices, or nodes, each representing a possible event (a so-called random variable), and a set of edges, or links, representing the dependency relationships between these events. Such graphs provide an intuitive, visual method for understanding and manipulating relationships between events that are often obscured by algebraic or tabular representations. PGMs come in a variety of forms, such as Bayesian (so-called “belief”) networks, and are similar to other kinds of graphical models, such as decision graphs and factor graphs. Here I will combine these to produce a clear representation of the essential structure of the problem.

Visually, we may represent the random variables that constitute the Monty Hall problem, and their dependent relationships, as a PGM in Figure \ref{fig:monty-hall-graph}. In this diagram we adopt the conventions of a decision graph, using squares to represent decisions, circles to represent purely chance events, and the triangle to represent a utility function that is also the end of the story. In addition,  we use shaded nodes to represent those enacted by the host and unshaded nodes to represent those of the contestant. The sequence represented by this graph consists of the following events, in order of their occurrence, each associated with a function and set of outcomes described below:

\begin{figure*}
    \centering
    \includegraphics[width=1\linewidth]{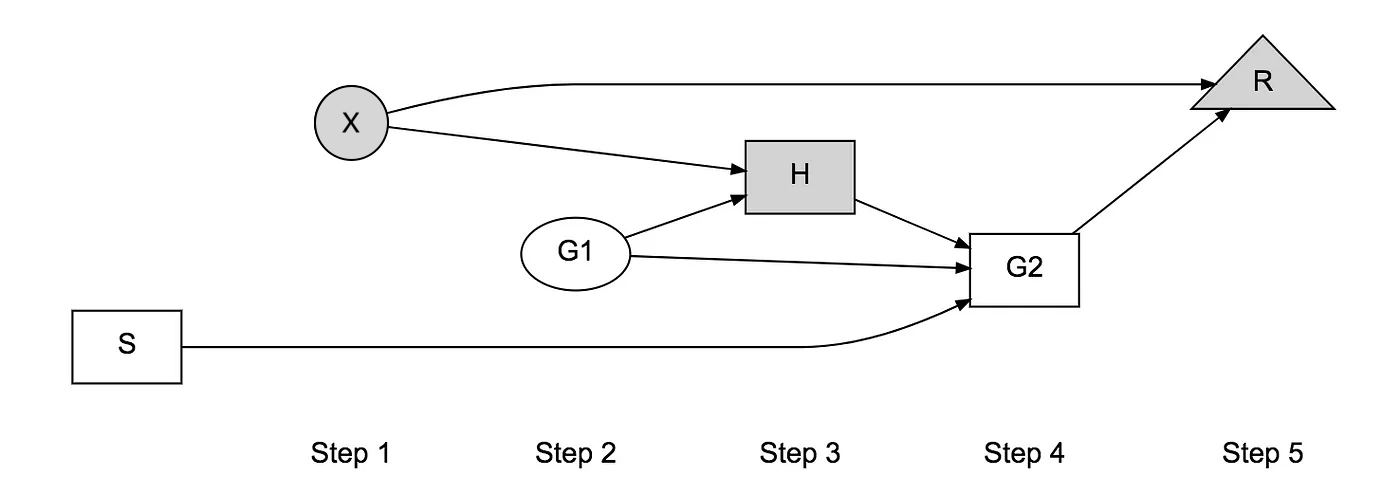}
    \caption{A graphical model of the Monty Hall Problem}
    \label{fig:monty-hall-graph}
\end{figure*}

\begin{itemize}
\item[] $S$:  The strategy employed by the contestant.
\item[] $X$:  The door behind which the car has been placed.
\item[] $G_1$: The door guessed first by the contestant.
\item[] $H$:  The door revealed by the host.
\item[] $G_2$: The door guessed second by the contestant. May be same as $G_1$.
\item[] $R$:  The result of the contestant's second guess.
\end{itemize}

The nodes in this graph are essentially the columns in our first table, with the addition of the random variable $S$, which will stand for the strategy adopted by the contestant. This event is rarely, if ever, mentioned in the telling of the problem, but it is helpful in illuminating its overall structure. The contestant actually has three possible strategies—(1) she may keep her first guess, or (2) she may switch to the door offered by the host after he reveals a goat, or—and this is rarely mentioned but actually assumed by the incorrect solution—(3) she may randomly choose, say by flipping a coin, whether to switch or not. Thus, we can define $S$ as having three outcomes: $\{keep, switch, flip\}$. (Note that these values are actually segments of a continuous weight value applied to the switching (or keeping) strategy, where $0$ means $keep$ and $1$ means $switch$). What’s interesting about this event is that it actually precedes the game—in fact, the function refers to what you are doing right now by reading this essay and following its argument, and any time the problem is discussed. We are figuring out which strategy to choose and returning a value as our answer.

As an aside, we could also model the policy of the host with another random variable, say $P$, on which $H$ would depend. This would reflect the host’s role as either ‘good host’ or ‘bad host,’ or as some intermediate state. The good host wants the contestant to win, and so would always show the goat when the contestant has guessed incorrectly, and would never show the goat when the contestant guesses correctly. The bad host would do the opposite. Like $S$, the outcomes of $P$ may be represented as three states, $\{good, bad, neutral\}$, even though they form a continuous weighting factor between $0$ and $1$ applied to the strategy of being a good host. However, we do not include this event in our model because, again, its existence would undermine the integrity of the question.

\subsection{Elements of the graph}

Let’s go over each of the elements of the graph diagram. We will assume that the doors mentioned in each event are elements of the set $D = \{A, B, C\}$ and that each random variable has access to it; so, we don’t represent $D$ as a node. Also, we’ll use uppercase letters to stand for the random variables, and their lowercase variants to stand for their outcomes in a specific trial, or play of the game. Finally, in each case we will describe the function along with its possible outcomes and associated probabilities using tables. These tables describe “local probabilities”—the probability of each outcome from a function considered relative to the total set of outcomes for that random variable.

\textbf{The first random variable} $X$ is the selection of the door behind which the car has been placed, and by implication, it entails the doors behind which the goats have been placed.  We assume that the selection is a purely random event, so its function and possible outcomes are those in Table \ref{tab:outcomes-for-c}.
$$
X(): x \in D
$$
\begin{table}
    \centering
    \begin{tabular}{ccc}
        \toprule
         A& B &C\\
         \midrule
         .33&  .33&.33\\
        \bottomrule
    \end{tabular}
    \vspace{5px}
    \caption{Outcomes for X}
    \label{tab:outcomes-for-c}
\end{table}

\textbf{The second random variable} $G_1$ is the door the contestant first guesses. It too is a purely random event, since the contestant has no information to go on. So, its function and outcomes are an equiprobable distribution over $D$ (see Table \ref{tab:outcomes-for-g1}).
$$
G_1(): g_1 \in D
$$
\begin{table}
    \centering
    \begin{tabular}{ccc}
        \toprule
         A& B &C\\
         \midrule
         .33&  .33&.33\\
        \bottomrule
    \end{tabular}
    \vspace{5px}
    \caption{Outcomes for $G_1$}
    \label{tab:outcomes-for-g1}
\end{table}

\textbf{The third random variable} $H$ is the door selected by the host. Unlike the first two, it is not entirely random and depends on both $x$ and $g_1$ as inputs. In the language of decision graphs, it is a decision, as opposed to the first two, which are chance events. $H$ can be specified as a set function:
$$
H(x, g_1): h \in D - \{x \cup g_1\}
$$
This function succinctly describes the action of the host in revealing a door. It returns one member of the set produced by subtracting from the total set of doors, $D$, the set formed by the union of the door hiding the car and the door guessed by the contestant. When $c$ and $g_1$ are identical, the result will be two doors to choose from, and H returns a random selection among them; when they are different, $H$ can return only one value. Note that it does not matter if the host is biased in his selection of a door in the case when he has a choice, as either will have a goat behind it. In fact, we could replace the random function with a fixed rule, to choose the first element of the set, and we would get the same result. The outcomes are represented in Table \ref{tab:outcomes-for-h}.

\begin{table}
    \centering
    \begin{tabular}{ccccc}
        \toprule
         $x$& $g_1$& $A$& $B$& $C$\\
         \midrule
         A& A& 0& .5& .5\\
         A& B& 0& 0& 1\\
         A& C& 0& 1& 0\\
         \midrule
         B& A& 0& 0& 1\\
         B& B& .5& 0& .5\\
         B& C& 1& 0& 0\\
         \midrule
         C& A& 0& 1& 0\\
         C& B& 1& 0& 0\\
         C& C& .5& .5& 0\\         
        \bottomrule
    \end{tabular}
    \vspace{5px}
    \caption{Outcomes of $H$ given $c$ and $g_1$}
    \label{tab:outcomes-for-h}
\end{table}

\textbf{The fourth random variable} $G_2$ is the contestant’s second guess. It is also a decision, and it depends on the results of three inputs, $G_1$, $H$, and $S$. The function $G_2$ looks like this:

\[ G_2(g_1, h, s) : \left\{ \begin{array}{ll}
        s = switch: & g_2 \in D - \{h \cup g_1\} \\
        s = keep: & g_2 = g_1 \\
        s = flip: & g_2 \in D - h
\end{array} \right. \]

The first case is a set function similar to that of $H$, applied to the values produced by $X$ and $G_1$. In the second case we simply to keep the value produced by $G_1$. In the third case, we randomly choose between the two values provided by the previous cases. We can represent the outcomes as three tables (\ref{tab:outcomes-for-g2-switch}, \ref{tab:outcomes-for-g2-keep}, and \ref{tab:outcomes-for-g2-flip}), one for each strategy.

\begin{table}
    \centering
    \begin{tabular}{ccccc}
        \toprule
         $g_1$& $h$& $A$& $B$& $C$\\
         \midrule
         A& B& 0& 0& 1\\
         A& C& 0& 1& 0\\
         \midrule
         B& A& 0& 0& 1\\
         B& C& 1& 0& 0\\
         \midrule
         C& A& 0& 1& 0\\
         C& B& 1& 0& 0\\         
        \bottomrule
    \end{tabular}
    \vspace{5px}
    \caption{Outcomes for $G_2$, given the “switch” strategy, $g_1$, and $h$}
    \label{tab:outcomes-for-g2-switch}
\end{table}

\begin{table}
    \centering
    \begin{tabular}{ccccc}
        \toprule
         $g_1$& $h$& $A$& $B$& $C$\\
         \midrule
         A& B& 1& 0& 0\\
         A& C& 1& 0& 0\\
         \midrule
         B& A& 0& 1& 0\\
         B& C& 0& 1& 0\\
         \midrule
         C& A& 0& 0& 1\\
         C& B& 0& 0& 1\\         
        \bottomrule
    \end{tabular}
    \vspace{5px}
    \caption{Outcomes for $G_2$, given the “keep” strategy, $g_1$, and $h$}
    \label{tab:outcomes-for-g2-keep}
\end{table}

\begin{table}
    \centering
    \begin{tabular}{ccccc}
        \toprule
         $g_1$& $h$& $A$& $B$& $C$\\
         \midrule
         A& B& .5& 0& .5\\
         A& C& .5& .5& 0\\
         \midrule         B& A& 0& .5& .5\\
         B& C& .5& .5& 0\\
         \midrule
         C& A& 0& .5& .5\\
         C& B& .5& 0& .5\\         
        \bottomrule
    \end{tabular}
    \vspace{5px}
    \caption{Outcomes for $G_2$, given the “flip” strategy, $g_1$, and $h$}
    \label{tab:outcomes-for-g2-flip}
\end{table}

\textbf{The final random variable} $R$ is the result of the game, which simply compares the results of $g_2$ and $x$ for equality and returns a value of $true$ or $false$, which stand for winning and losing respectively. In the language of decision graphs, this is a utility function with outcomes represented in Table \ref{tab:outcomes-for-r} and the following form:

$$
R(x, g_2): r = x \equiv g_2
$$

\begin{table}
    \centering
    \begin{tabular}{ccccc}
        \toprule
         $x$& $g_2$& $T$& $F$ \\
         \midrule
         A& A& 1& 0 \\
         A& B& 0& 1 \\
         A& C& 0& 1 \\
         \midrule
         B& A& 0& 1 \\
         B& B& 1& 0 \\
         B& C& 0& 1 \\
         \midrule
         C& A& 0& 1 \\
         C& B& 0& 1 \\
         C& C& 1& 0 \\         
        \bottomrule
    \end{tabular}
    \vspace{5px}
    \caption{Outcomes for $R$, given $x$ and $g_2$}
    \label{tab:outcomes-for-r}
\end{table}

\subsection{Modeling Strategy}

Note that we have not described a function for the random variable $S$, nor have we associated its outcomes with any probabilities. This is because, unlike the other functions, $S$ entails a complex set of conditions and functions that, as mentioned above, include this essay and any discourse relating to the problem. It is effectively a black box for the purposes of our representation.  Moreover, its outcome is a normative fact, not a descriptive one---we are deciding upon a rule that will govern the behavior of any future contestant.

One immediate advantage to the preceding graphical representation is that it focuses our attention on the two decision events, $H$ and $G_2$, which are essential to providing an intuitive understanding of the problem. To describe these is, in effect, to understand the problem. As we have seen, there are two salient facts to consider in this regard. First, the host is constrained to the point of essentially giving away the answer two thirds of the time. Second, the contestant may employ at least three strategies in making her decision to switch or not. Implicitly, the nature of the problem is to infer the structure of $H$ in order to establish a rule for $G_2$.

Let us now consider the three strategies available to the contestant. We have seen that a strategy to $keep$ results in winning $\frac{1}{3}$ of the time, whereas the strategy to $switch$ results in winning $\frac{2}{3}$ of the time. We have also noted that there is another strategy---to $flip$, or randomly choose whether to switch or not. (One may imagine the contestant flipping a coin here.) To determine the result of this strategy, and to demonstrate the value of the PGM approach, we generate a set of decision trees from our graph, which may be interpreted as a generative model, and the functions just described.

\begin{figure}[ht]
    \centering
    \includegraphics[width=0.9\linewidth]{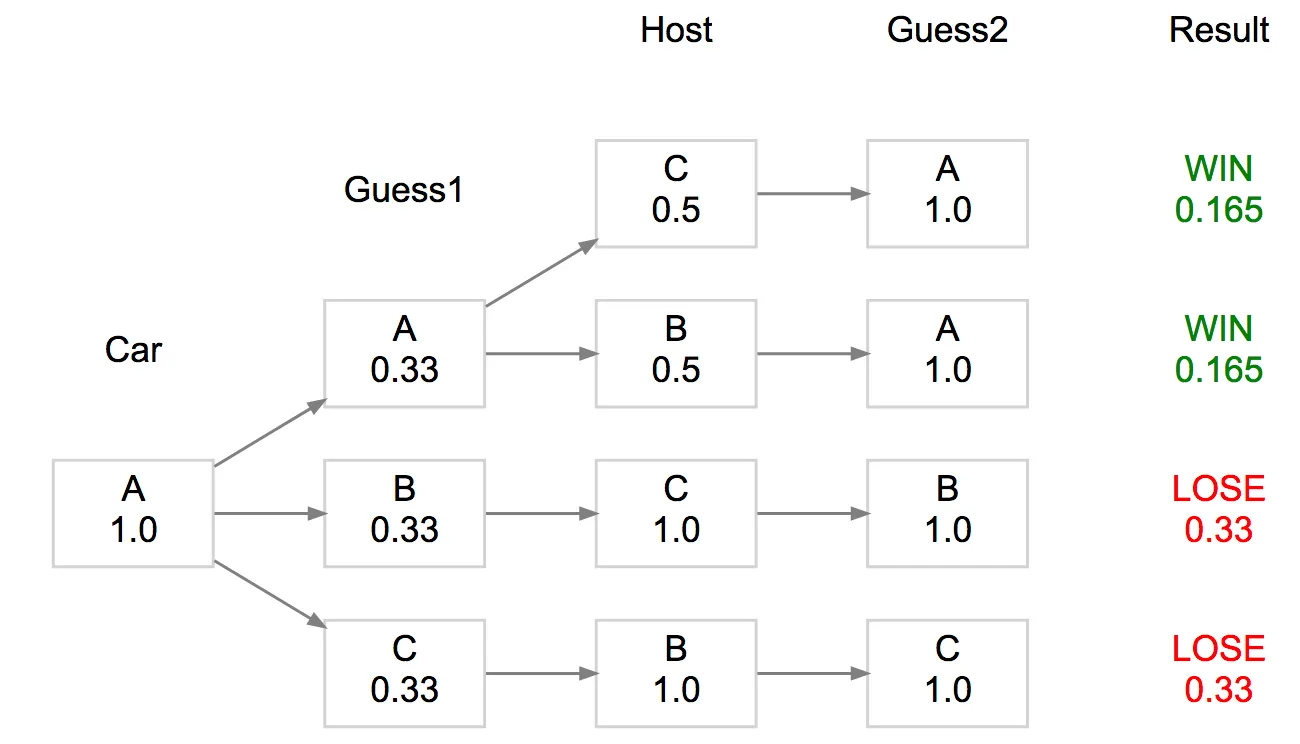}
    \caption{Decision tree for $keeping$ strategy}
    \label{fig:tree-keeping}
\end{figure}

\begin{figure}[ht]
    \centering
    \includegraphics[width=0.9\linewidth]{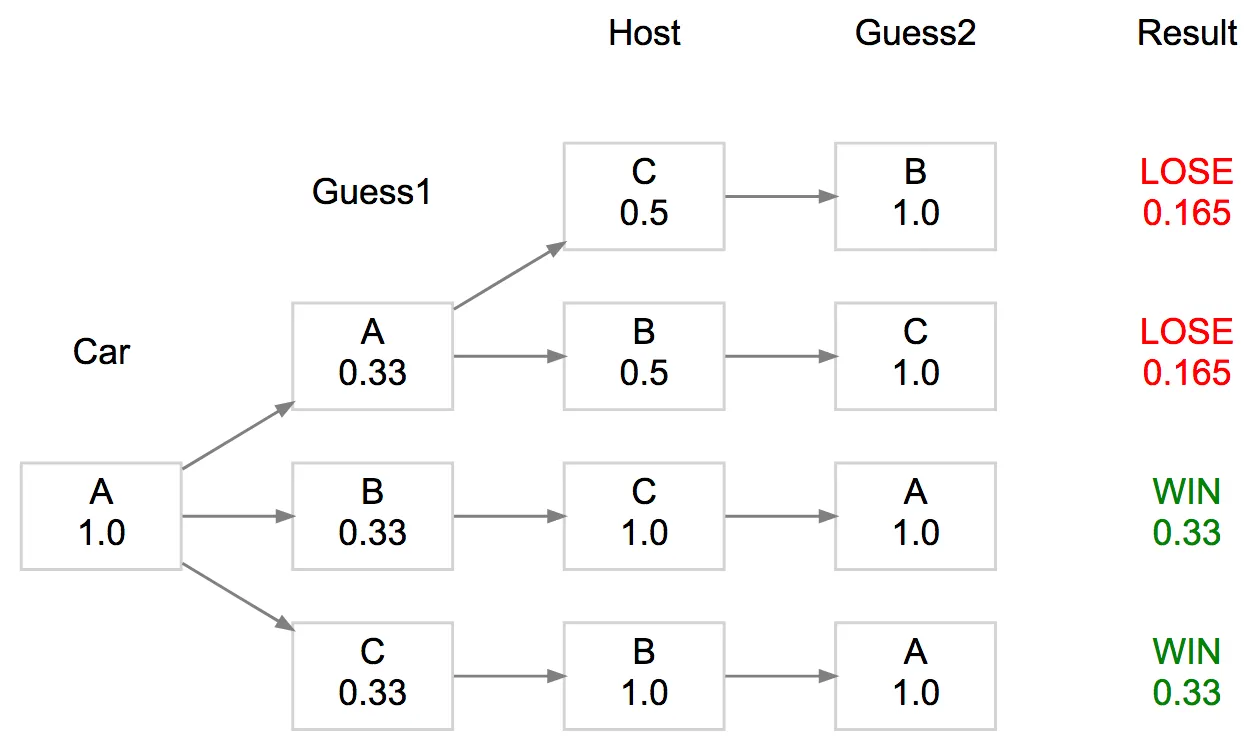}
    \caption{Decision tree for $switching$ strategy}
    \label{fig:tree-switching}
\end{figure}

\begin{figure}[ht]
    \centering
    \includegraphics[width=0.9\linewidth]{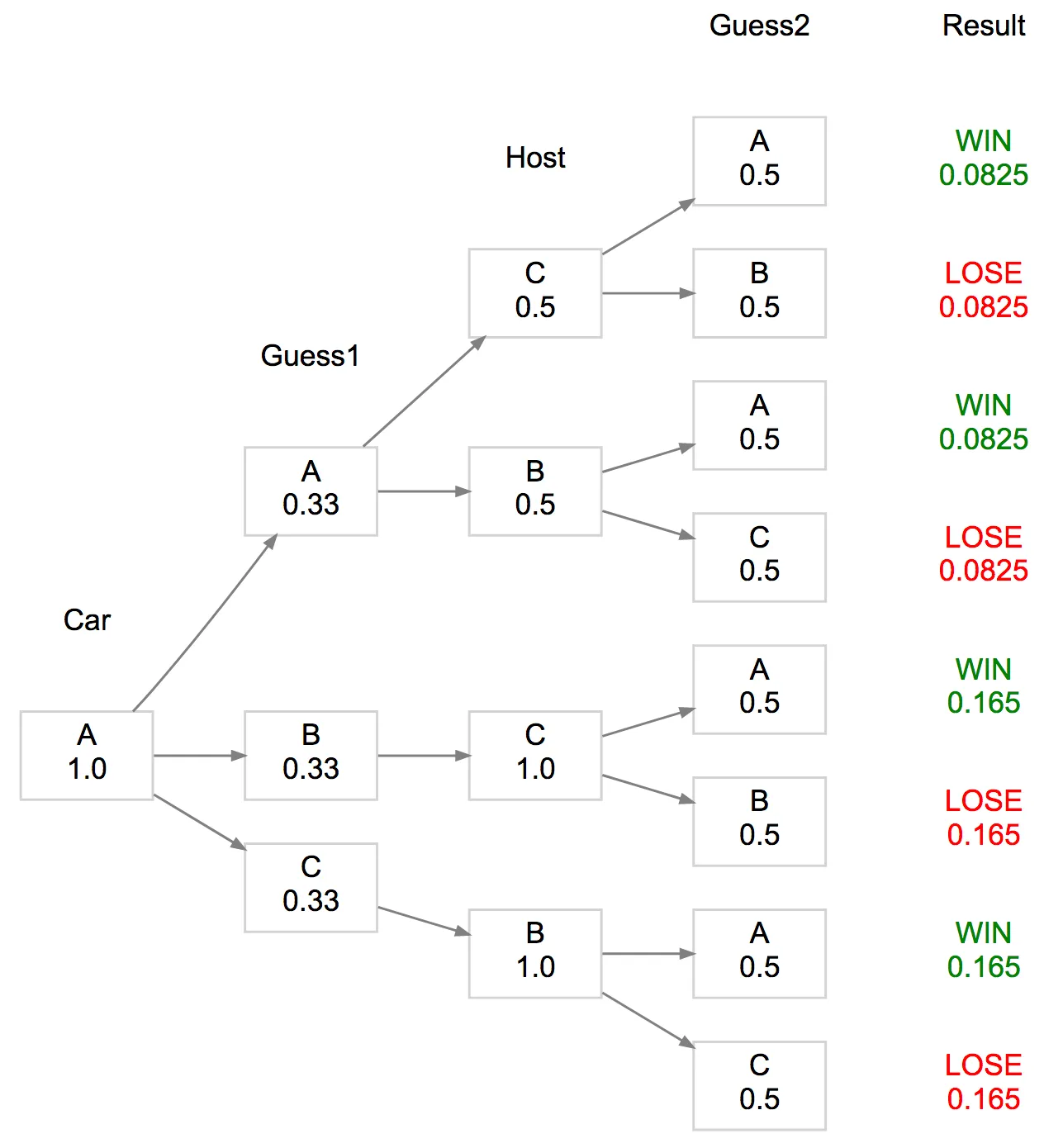}
    \caption{Decision tree for $flipping$ strategy}
    \label{fig:tree-flipping}
\end{figure}

Figures \ref{fig:tree-keeping}, \ref{fig:tree-switching}, and \ref{fig:tree-flipping} show the generated trees fore each of the strategies. In each case, the tree is restricted to the case where the car is behind door $A$, since the results are identical for each placement of the car. In addition, the rectangles stand for the specific outcomes and local probabilities of each event function, which are given in the column names. Finally, the results of each outcome are computed by the product of the ancestors of each. For example, the probability for the first outcome in the first diagram, $AACA$, can be computed as $1.0 \times 0.33 \times 0.5 \times 1.0 = 0.165$.

Table \ref{tab:summary-of-strategy} shows a summary of the results of each strategy. A  surprising outcome from this exercise is that it demonstrates the sense in which the incorrect solution may be viewed as correct. We see that the odds of winning really are $50:50$ if the contestant \textit{behaves} randomly when choosing among the two closed doors. But this is not how the answer is typically, if ever, given. Instead, the potential randomness of the contestant's behavior is objectified and projected onto the structure of the game itself. Call it misplaced randomness.

\begin{table}[ht]
    \centering
    \begin{tabular}{lcc}
        \toprule
          $s$& $W$ & $L$ \\
         \midrule
         $keep$  & .33 & .66 \\
         $switch$ & .66 & .33 \\
         $flip$  & .50 & .50 \\
        \bottomrule
    \end{tabular}
    \vspace{5px}
    \caption{Summary of results for each strategy}
    \label{tab:summary-of-strategy}
\end{table}




\section{Conclusion}

PGMs, as representations of the dependency structure of events in the problem, provide a useful means of clarifying the logic of the Monty Hall problem. They help us make sense of why the solution to the puzzle strikes us as hard to believe in the first place: the difficulty resides from the fact that although each successful approach is able to marshal the relevant data into an inductive format or deductive formula that produces the correct result, none of them formally represent the salient structure that actually generates the result. By isolating and connecting the elementary random variables that compose the problem, and by requiring us to specify the actual functions that generate the outcomes of each random variable, we see that the solution follows from  the latent structure of these functions. We end up realizing that the correct answer to the problem is not so much that the contestant will win $\frac{2}{3}$ of the time by switching, so much as that the car is behind the remaining door $\frac{2}{3}$ of the time.  We have discovered an objective condition to which one’s behavior should adapt. Strategy follows structure.

A larger lesson is that the problem strikes us as hard because we tend to frame probability problems primarily in terms of independence relations, and to consider structural relations, i.e. dependencies, as secondary. Only after some coaxing do we overcome this bias. Although many are quick to attribute this bias to a just-so story about human evolution, a more plausible explanation for it is our schooling and culture. The default mode of teaching probability is to cover independence first and dependence, in the form of conditional probability, second.  This makes sense from the perspective of starting with the simpler case, since it's easier to teach how to compute the joint probability of independent events $p(x,y) = p(x)p(y)$ than dependent ones $p(x,y) = p(x)p(y|x)$. In the first case, you don't have to explain what "$y$ given $x$" means. But simplicity comes at the cost of understanding, since the independent version is a special case of the dependent one. Foregrounding the former obscures the fact that the dependent case is the prior case---that, in fact, joint probability presupposes conditional probability. We miss the opportunity to show that the idea of a condition is essentially that of relationship, an element of a model that is implied by and motivates the algebraic formulae that we foreground in our teaching. 

As Pearl argued long ago (1988) \cite{pearlProbabilisticReasoningIntelligent1988}, independence buys us a great deal computationally, since it allows us to calculate reasonably accurate answers for a variety of complex problems. But it does so at the cost of semantic coherency---the idea that a solution should not only fit the probabilistic patterns observed in data but also align with meaningful and plausible relationships in the real world. In the case of the Monty Hall problem, what makes it hard is that the relationships that constitute the game---the dependency structure between events that can be modeled by a PGM---are rarely made explicit. The result is that although we may grasp the logic of the math and come around to accepting the correct solution, we may still wonder why the math works. By representing the problem as a structured series of events, this difficulty is removed.\footnote{For a computational representation of the model presented here, see the GitHub repo \url{https://github.com/ontoligent/montyhall}.}

\section{Addendum}

I owe to my colleague Pete Alonzi (UVA School of Data Science) the most elegant solution to the problem. In this approach, a sample space is generated from the two scenarios---whether the initial guess $g_1$ is correct---and two strategies---$keep$ or $switch$. Their combination yields four disjoint and exhaustive events, each of which may be assigned a joint probability which must sum to one. 

Referring to Table \ref{tab:alonzi-solution}, the probabilities are deduced in the following way. The probability that the contestant is correct the first time and does not switch is obviously $\frac{1}{3}$, since this is just the probability of guessing correctly the first time. Now, it is also obvious that the probability of guessing correctly the first time and switching is $0$. Similarly, the probability of guessing \textit{incorrectly} the first time and \textit{not} switching is $0$. This leaves $\frac{2}{3}$ for the final event, of being incorrect and switching---the event in question---since the sum of the probabilities of all the events must be $1$.

\begin{table}
    \centering
    \begin{tabular}{ccc}
    \toprule
         $g_1 \equiv x$  & $s = switch$ & $p(g_1, s)$\\
        \midrule
         1 & 0 & $\frac{1}{3}$\\
         1 & 1 & 0 \\
         0 & 0 & 0 \\
         0 & 1 & $\frac{2}{3}$ \\
         \midrule
         & SUM & 1 \\
         \bottomrule
    \end{tabular}
    \vspace{5px} 
    \caption{Joint probabilities of scenarios and strategies}
    \label{tab:alonzi-solution}
\end{table}

This solution is highly compact and requires very few assumptions. The main assumption is in the sample space selection---instead of focusing on the specific events in the game play, it focuses on the two scenarios identified earlier in this paper, which essentially group the more primitive events in Table \ref{tab:sample-space-01}. The trick to grasping this solution is in accepting that these scenarios are in fact events.

With respect to the thesis of this paper, that the Monty Hall problem is hard because of the fore-grounding of the independence assumption, Alonzi's solution has the virtue of by-passing the issue by focusing on the deduction of joint probabilities. However, it does suffer from a certain opacity precisely due to its simplicity. Although its logic is ironclad, its connection to the specifics of the game remain unclear. For example, it seems to be unconnected to such parameters as how many doors the game involves. In fact, the solution appears to generalize over any number of doors.

To explore how it generalizes, consider the joint probabilities when we adjust the number of doors $|D|$ to $4$ as shown in Table \ref{tab:alonzi-solution-4doors}. We find that the event of being incorrect and switching also has a higher probability $\frac{3}{4}$ than the other events. But the question becomes, what are we in fact selecting in this event when $|D| > 3$? The probability in this case refers to any one of the remaining two doors. To compute the probability of just one of these doors, however, requires the introduction of a conditional probability: the probability of one of the doors having the car given the event, i.e. $p(g_2|g_1 \equiv x, s = switch; |D| = 4)$. This turns out to be $\frac{3}{4} \times \frac{1}{2} = \frac{3}{8}$. This value is $> \frac{1}{4}$ or, perhaps more clearly, $\frac{2}{8}$.

\begin{table}
    \centering
    \begin{tabular}{ccc}
    \toprule
         $g_1 \equiv x$  & $s = switch$ & $p(g_1, s; |D| = 4)$\\
        \midrule
         1 & 0 & $\frac{1}{4}$\\
         1 & 1 & 0 \\
         0 & 0 & 0 \\
         0 & 1 & $\frac{3}{4}$ \\
         \midrule
         & SUM & 1 \\
         \bottomrule
    \end{tabular}
    \vspace{5px} 
    \caption{Joint probabilities of scenarios and strategies for 4 doors}
    \label{tab:alonzi-solution-4doors}
\end{table}
 Now, we can generalize this formula and see that no matter how many doors there are in the game, the decision to switch will always be greater than not switching, although as $|D|$ increases, the difference will become less significant. If we let $n$ stand for $|D|$, we get the following formula for the $(0, 1)$ event:

 $$
\frac{n-1}{n} \times \frac{1}{n - 2}
$$

Which reduces to:

$$
\frac{n-1}{n^2 - 2n}
$$

\begin{figure}
    \centering
    \includegraphics[width=1\linewidth]{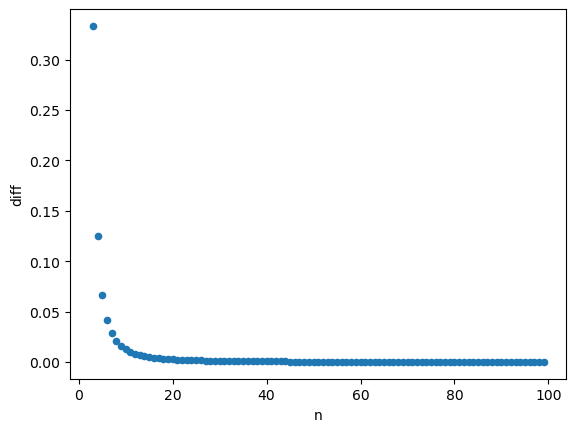}
    \caption{Probability difference of switching vs keeping as $n$ increases}
    \label{fig:n-doors}
\end{figure}

This value will always be $> \frac{1}{n}$, although the difference between the choices declines precipitously as $n$ increases (see Figure \ref{fig:n-doors}).

When we apply the general formula back to our specific case, we find:

$$
\frac{3-1}{3^2 - 2 \times 3} = \frac{2}{9 - 6} = \frac{2}{3}
$$

Alonzi's solution elegantly illustrates the power of a probabilistic solution coupled with a judiciously selected sample space. By choosing the right sample space, the solution becomes both clear and simple, by-passing the need to apply the more complex Bayesian formula. Perhaps the same principle applies to math as was expressed by Rob Pike many years ago for programming (1989) \cite{pike1989notes}:

\begin{quote}
Data dominates. If you’ve chosen the right data structures and organized things well, the algorithms will almost always be self-evident. Data structures, not algorithms, are central to programming.    
\end{quote}

If we think of data structures more broadly to include dependency structures, then the thesis of this essay may be considered a contribution to this school of thought.

\printbibliography

\end{document}